\documentclass[twocolumn,showpacs,preprintnumbers,amsmath,amssymb]{revtex4}

\newcommand{\bib}{\bibitem}
\newcommand{\bea}{\begin{eqnarray}}
\newcommand{\eea}{\end{eqnarray}}
\newcommand{\beq}{\begin{equation}}
\newcommand{\eeq}{\end{equation}}
\newcommand{\non}{\nonumber}
\newcommand{\noi}{\noindent}

\newcommand{\de}{\delta}
\newcommand{\ep}{\epsilon}

\newcommand{\la}{\lambda}

\newcommand{\si}{\sigma}
\newcommand{\pa}{\partial}
\newcommand{\bom}{\bf m}
\newcommand{\bn}{\bf n}
\newcommand{\bx}{\bf x}
\newcommand{\by}{\bf y}
\newcommand{\bz}{\bf z}

\usepackage{graphicx,epsfig} % Include figure files
\usepackage{dcolumn} % Align table columns on decimal point
\usepackage{bm} % bold math

\begin{document}

\title{Spin-$S$ Kitaev model: \\
Classical Ground States, Order from Disorder \& Exact Correlation Functions}
\author{G. Baskaran$^1$, Diptiman Sen$^2$ and R. Shankar$^1$}
\affiliation{$^1$The Institute of Mathematical Sciences, CIT Campus, 
Chennai 600 113, India \\ $^2$Center for High Energy
Physics, Indian Institute of Science, Bangalore, 560 012, India}
\date{\today}

\begin{abstract} 
In the first part of this paper, we study the spin-$S$ Kitaev model using 
spin wave theory. We discover a remarkable geometry of the minimum energy 
surface in the $N$-spin space. The classical ground states, called Cartesian 
or CN-ground states, whose number grows exponentially 
with the number of spins $N$, form a set of points in the $N$-spin space. 
These points are connected by a network of {\it flat valleys} in the 
$N$-spin space, giving rise to a continuous family of classical ground
states. Further, the CN-ground states have a correspondence with {\it dimer 
coverings} and with {\it self avoiding walks} on 
a honeycomb lattice. The zero point energy of our spin wave theory picks 
out a subset from a continuous family of classically degenerate states as 
the quantum ground states; the number of these states also grows exponentially
with $N$. In the second part, we present some exact results. For arbitrary 
spin-$S$, we show that localized $Z_2$ flux excitations are present by 
constructing plaquette operators with eigenvalues $\pm 1$ which commute with 
the Hamiltonian. This set of commuting plaquette operators leads to an exact 
vanishing of the spin-spin correlation functions, beyond nearest neighbor 
separation, found earlier for the spin-1/2 model \cite{baskaran1}.
We introduce a generalized Jordan-Wigner 
transformation for the case of general spin-$S$, and find a 
complete set of commuting link operators, similar to the spin-1/2 model, 
thereby making the $Z_2$ gauge structure more manifest. The Jordan-Wigner 
construction also leads, in a natural fashion, to Majorana fermion operators
for half-odd-integer spin cases and hard-core boson operators for integer spin
cases, strongly suggesting the presence of Majorana fermion and boson 
excitations in the respective low energy sectors. Finally, we present a 
modified Kitaev Hamiltonian which is exactly solvable for all half-odd-integer
spins; it is equivalent to an exponentially large number of copies of 
spin-1/2 Kitaev Hamiltonians.
\end{abstract}

\maketitle

\section{Introduction} 

Frustrated quantum spin systems have become a new paradigm in condensed
matter science. More and more systems are joining this family. A richness is 
emerging in terms of novel ground states and excitations. They not only 
enrich the basic science of strongly correlated electron systems, but also 
have started playing a key role in an unexpected corner, namely quantum 
computers. Kitaev has suggested that excitations of frustrated quantum spin 
systems have a special robustness, arising from their non-trivial topological
property, which make them suitable elements of a topological quantum computer.
An exactly solvable two-dimensional frustrated spin-1/2 model introduced by 
Kitaev \cite{kitaev1,kitaev2} exemplifies this. 

The spin-1/2 Kitaev model is interesting in its own right as a condensed matter
spin model \cite{feng,nussinov1,nussinov2,baskaran1,vidal,lee,sengupta}. 
In fact a similar model, called the {\it compass model}, although 
not exactly solvable, was introduced by Kugel and Khomskii in the late
1970s \cite{kugel} to understand the magnetic properties of transition metal 
oxides which have orbital degeneracies. Recently an optical lattice 
realization of the spin-1/2 Kitaev model has been discussed \cite{demler}.

The spin-$S$ Kitaev model for $S > 1/2$ is not exactly solvable. It 
is a challenging question if the $Z_2$ gauge structure and the presence of 
low energy Majorana fermions, discovered by Kitaev for the spin-1/2 case, 
survive for arbitrary spin-$S$. Are there differences between half-odd-integer
and integer spins? In the present paper we approach the problem on two fronts.
First we find exact classical ground states and perform a spin wave analysis. 
The structure of the ground state manifold in the $N$-spin space is rich: an 
exponentially large number of isolated points are connected by flat valleys. 
We will call the ground state corresponding to these isolated points as 
Cartesian or CN-ground states, as any given spin points along one of the 
three Cartesian directions. We also find the phenomenon of order from disorder
in our spin wave analysis. We discover a nice connection between finding the 
CN-ground states and the dimer covering problem on the honeycomb lattice. In 
our spin wave analysis we find an equal number of finite frequency and zero 
frequency spin wave modes which live on self avoiding walks that are uniquely 
connected to the dimer covering.

On the other front, we get some useful and exact results for the spin-$S$ 
Kitaev model, which prove the survival of the $Z_2$ gauge structure. There 
are also good indications that the low energy Majorana fermion excitations 
survive for half-odd-integer spins. Specifically we find plaquette operators 
(with eigenvalues $\pm 1$) which commute with the Hamiltonian. This set of 
commuting plaquette operators leads to a vanishing of the spin-spin 
correlation functions, beyond nearest neighbor separation, found earlier 
for the spin-1/2 model \cite{baskaran1}. We also discover a Jordan-Wigner 
transformation for arbitrary $S$ which leads to new bond operators (with 
eigenvalues $\pm 1$) which commute with the Hamiltonian. This makes the $Z_2$
gauge structure manifest. The Jordan-Wigner construction leads, in a natural 
fashion, to Majorana fermion operators for the case of half-odd-integer 
spins and to hard-core boson operators for the integer spin case, strongly 
suggesting the presence of Majorana fermions and bosons in the respective 
low energy sectors.

The plan of the paper is as follows. In Sec. II, we consider a one-dimensional
version of the Kitaev model and study spin wave theory in the large $S$ limit.
We find that although there is a continuous family of classical ground states
parametrized by an angle, the zero point energy of the spin waves picks
out a discrete set of values of the angle as the quantum ground states; these
correspond to dimer coverings on alternate bonds of the model. In Sec. III,
we study the two-dimensional Kitaev model using spin wave theory. We first 
present a general argument to find the classical ground state energy. We then 
identify a discrete and infinite number of classical ground states; these have
an interesting correspondence with self avoiding walks (SAW) and dimer 
coverings of the honeycomb lattice. The spin wave spectrum is found to 
contain {\it one-dimensional} finite frequency spin wave modes and an equal 
number of zero frequency modes living on the SAWs.
The zero point energy of the spin waves again picks 
out a special class of self avoiding walks as the quantum ground states.
The number of classical and quantum ground states both grow exponentially with
the number of sites, although the latter number grows slower than the 
former. In Sec. IV, we construct, for any value of the spin $S$, an infinite 
set of $Z_2$ operators which commute with each other and with the Hamiltonian.
We use these operators to show that in any eigenstate of the Hamiltonian, the 
spin-spin correlations vanish unless the two spins are nearest 
neighbors; even for nearest neighbors, only certain components of the 
correlations are non-zero. Finally, we use a Jordan-Wigner-like transformation
to construct a set of operators which act like Majorana fermions (hard-core 
bosons) for half-odd-integer (integer) values of $S$ respectively. The $Z_2$
operators defined earlier can be written as products of the Majorana fermion
(hard-core boson) operators. In Sec. V, we present a modified Kitaev 
Hamiltonian whose energy spectrum can be found for any value of 
half-odd-integer spin; this model is equivalent to an exponentially large
number of copies of the spin-1/2 Kitaev Hamiltonian. Some directions for 
future work are pointed out in Sec. VI.

\section{One-dimensional Kitaev model}

In this section, we will discuss a one-dimensional spin-$S$ model which is 
obtained by considering a single row of the Kitaev model in two dimensions.
We illustrate the order from disorder phenomenon explicitly. The model is a 
spin $S$ chain governed by the Hamiltonian
\beq H_1 ~=~ \frac{J}{S} ~\sum_{i=-\infty}^\infty ~(S_{i,1}^x S_{i,2}^x ~+~ 
S_{i,2}^y S_{i+1,1}^y). \label{ham1} \eeq
We assume that $J > 0$. (If $J < 0$, we can change its sign by performing 
a unitary rotation which flips the signs of $S_{i,1}^x$, $S_{i,1}^z$,
$S_{i,2}^y$ and $S_{i,2}^z$ for all values of $i$). In Eq. (\ref{ham1}),
the unit cells are labeled by $i$, and each unit cell has two spins labeled
as 1 and 2. A factor of $1/S$ has been introduced in Eq. (\ref{ham1}) so that
the ground state energy is proportional to $S$ in the limit $S \to \infty$.

Let us introduce two vectors in the $x-y$ plane,
\bea \hat \bn &=& cos \theta ~\hat \bx ~+~sin \theta ~\hat \by \non \\
{\rm and} ~~\hat{\bf e} &=& - ~sin \theta ~\hat \bx ~+~ cos \theta ~ \hat 
\by. \eea
Then a classical ground state of the Hamiltonian in Eq. (\ref{ham1}) is given
by the configuration
\beq {\bf S}_{i,1}^{cl} ~=~ S ~\hat \bn ~~{\rm and} ~~{\bf S}_{i,2}^{cl} ~
=~ - ~S ~\hat \bn. \label{gnd} \eeq
The classical energy of this state is $E^{cl}=-JSN/2$, where $N$ is the number
of sites (the number of unit cells is $N/2$). Thus the classical ground states
form a continuous family parametrized by an angle $\theta$ which lies in the 
range $[0,2\pi]$. We will now perform a spin wave analysis and show that this 
picks out four values of $\theta$ as having the lowest zero point energy;
these correspond to Cartesian ground states.

The spin wave spectrum around the ground state given in Eq. (\ref{gnd})
can be found by 
using the Holstein-Primakoff (HP) transformation from spins to simple harmonic
oscillator raising and lowering operators \cite{holstein,anderson,kubo}. To 
obtain a HP Hamiltonian which is quadratic in bosons, we expand the fields as 
\bea {\bf S}_{i,1} &=& S \hat \bn \left( 1 -\frac{p_{i,1}^2 + q_{i,1}^2}{2S}
\right) +\sqrt{S} \left( \hat{\bf e}q_{i,1} + \hat \bz p_{i,1} \right), \non \\
{\bf S}_{i,2} &=& -S \hat \bn \left( 1 - \frac{p_{i,2}^2 + q_{i,2}^2}{2S}
\right) +\sqrt{S} \left( -\hat{\bf e}q_{i,2} +\hat \bz p_{i,2} \right), \non \\
& & \eea
where $[q_{i,a},p_{j,b}]=i\delta_{ij} \de_{ab}$.
The spin wave Hamiltonian is then given by
\bea & & H_{1,sw} \non \\
& & = J \sum_i ~( p_{i,1}^2 + q_{i,1}^2 + p_{i,2}^2 + q_{i,2}^2 \non \\
& & ~~~~~~~~~~~~- ~cos^2 \theta ~q_{i,1}q_{i,2} ~-~ sin^2 \theta ~q_{i,2} 
q_{i+1,1}) \non \\
& & = J \sum_{k=0}^\pi ~\left( \begin{array}{cc} p_{-k,1} & p_{-k,2}
\end{array} \right) \left( \begin{array}{cc} 1 & 0 \\ 
0 & 1 \end{array} \right) \left( \begin{array}{c} p_{k,1} \\
p_{k,2} \end{array} \right) \non \\
& & \non \\
& & ~~~~~+ \left( \begin{array}{cc} q_{-k,1} & q_{-k,2} \end{array} \right) 
\left( \begin{array}{cc} 1 & f(k) \\
f^*(k) & 1 \end{array} \right) \left( \begin{array}{c} q_{k,1}\\
q_{k,2} \end{array} \right), \non \\
{\rm where} & & f(k) ~\equiv~ - ~cos^2 \theta ~-~ sin^2 \theta ~e^{ik}, \eea
and $k$ goes in steps of $4\pi /N$. The spin wave energies are 
\bea \ep_{k+} &=& J \sqrt{1+\vert f(k)\vert}, \non \\
{\rm and} ~~\ep_{k-} &=& J \sqrt{1-\vert f(k) \vert}, \eea
where $\vert f(k) \vert = \sqrt{1 - sin^2 (2\theta)~ sin^2 (k/2)}$. The zero 
point energy is
\beq E_{1,sw} ~= ~J \sum_{k=0}^\pi \left( \sqrt{1 +\vert f(k) \vert} ~+~
\sqrt{1- \vert f(k) \vert} \right). \eeq
We see that for each value of $k$, the spin wave energies have the same values 
for $\theta$ and $\pi /2 - \theta$. Now,
\bea & & \frac{\pa}{\pa \vert f\vert } ~\left( \sqrt{1+ \vert f
\vert} ~+~ \sqrt{1- \vert f \vert} \right) \non \\
& & = ~\frac{1}{2} ~\left( \frac{1}{\sqrt{1+|f|}} ~-~ \frac{1}{\sqrt{1-|f|}}
\right) ~<~0. \eea
Thus the total spin wave energy $\ep_{k+}+\ep_{k-}$ increases monotonically 
as $\vert f\vert$ decreases, i.e., as $\theta$ increases from 0 to $\pi /4$ or
decreases from $\pi /2$ to $\pi /4$. Thus the zero point energy is minimum 
at $\theta=0, ~\pi /2, ~\pi$ and $3\pi /2$, thereby picking out four points 
from the continuous family of classical ground states. These four points 
correspond to all the spins pointing along the $\pm {\hat \bx}$ or $\pm 
{\hat \by}$ directions.

Interestingly, the ground states chosen by the {\it order from disorder} 
phenomenon have two degenerate and non-dispersing spin wave branches with
frequencies, $\ep_{k+} = \ep_{k-} = J$. It is easy to show that all these modes
are localized on nearest neighbor bonds that have zero interaction energy. 

Finally, let us briefly discuss the case in which the couplings are not equal 
on all the bonds. Suppose that the $xx$ couplings have a strength $J_x$ and 
the $yy$ couplings have a strength $J_y$. If $J_x > J_y$, we find that the 
classical ground states are given by states in which the spins 1 and 2 in
each unit cell point in the $+\hat \bx, - \hat \bx$ or $-\hat \bx, + \hat \bx$
directions. The classical ground state degeneracy is therefore $2^{N/2}$.
We find that this degeneracy is not broken by the zero point energy of the 
spin waves.

\section{Two-dimensional Kitaev model}

We will now consider the spin-$S$ Kitaev model in two dimensions. This is 
a model on a honeycomb lattice with the Hamiltonian 
\bea H_2 &=& \frac{J}{S} ~\sum_{j+l={\rm even}} ~(S_{j,l}^x S_{j+1,l}^x ~
+~ S_{j-1,l}^y S_{j,l}^y \non \\
& & ~~~~~~~~~~~~~~~~+~ S_{j,l}^z S_{j,l+1}^z), \label{ham2} \eea
where $j$ and $l$ denote the column and row indices of the honeycomb lattice,
respectively. We again assume, without loss of generality, that $J > 0$.
Note that each spin is coupled to three other spins through $xx$, $yy$ and 
$zz$ couplings; we will denote the corresponding bonds as $x$, $y$ and $z$ 
bonds, respectively. (We present a schematic picture of the model in Fig. 1).
We will first assume that the couplings on the three 
kinds of bonds are equal. Note that the honeycomb lattice is bipartite, with 
sites belonging to the two sublattices $A$ and $B$ having $j+l$ as even 
and odd, respectively. If the total number of sites $N$ is even, each
sublattice has $N/2$ sites.

\begin{figure}[htb] \rotatebox{0}{\includegraphics*[width=\linewidth]{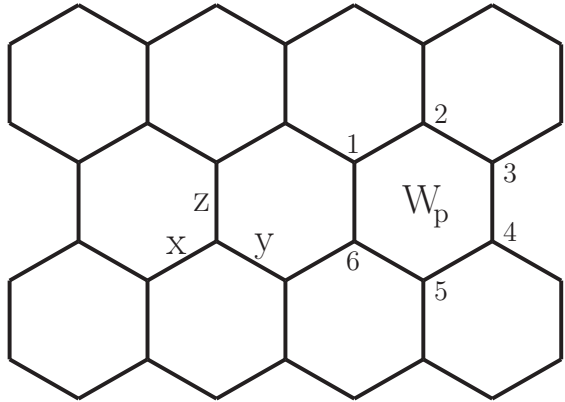}}
\caption{Schematic picture of the Kitaev model on a honeycomb lattice
indicating the three kinds of bonds $x$, $y$ and $z$. A hexagon with sites 
marked 1-6 is shown; the corresponding plaquette operator $W_p$ is defined
in Eqs. (\ref{fluxop}) and (\ref{fluxop12}).} \label{fig1} \end{figure}

We first present a general argument to obtain the classical ground state
energy of the Hamiltonian in Eq. (\ref{ham2}). We consider the spins at
different sites ${\bf S}_{\bn}$ to be classical (commuting) vectors,
and introduce a Lagrange multiplier $\la_{\bn}$ at each site to enforce the 
relation ${\bf S}_{\bn}^2 = S^2$; we do this by adding a term
\beq H_{\la} ~=~ -~ \frac{J}{2S} ~\sum_{\bn} ~\la_{\bn} ~({\bf S}_{\bn}^2 ~
-~ S^2) \label{hla} \eeq
to Eq. (\ref{ham2}). Extremizing the sum of Eqs. (\ref{ham2}-\ref{hla}) 
leads to the equations
\bea S_{\bn + \bom}^a &=& \la_{\bn} S_{\bn}^a , \non \\
S_{\bn}^a &=& \la_{\bn + \bom} S_{\bn + \bom}^a , \label{las} \eea
for any two neighboring sites $\bn$ and $\bn + \bom$ which are coupled by an 
$a$ bond ($a=x,y,z$). Substituting Eq. (\ref{las}) in Eq. (\ref{ham2}) and 
using the relation ${\bf S}_{\bn}^2 = S^2$, we find that the energy of such 
a state can be written in two ways which must be equal to each other, namely,
\beq E_{cl} ~=~ JS ~\sum_{\bn \in A} ~\la_{\bn} ~=~ JS ~\sum_{\bn \in B} ~
\la_{\bn}. \label{ecl} \eeq
Now, in any classical ground state, we can assume that for each site $\bn$, 
the spin on at least one of its three neighbors must point in such a 
direction that $S_{\bn}^a S_{\bn + \bom}^a \ne 0$. For such a pair, Eq. 
(\ref{las}) implies that $\la_{\bn} \la_{\bn + \bom} = 1$; note that
$\bn$ and $\bn + \bom$ necessarily belong to different sublattices. 
Extending this argument to all pairs of neighboring sites, we conclude that
$\la_{\bn}$ for all sites $\bn$ belonging to sublattice $A$ must have the
same value, denoted as $\la_A$, while $\la_{\bn}$ for all sites $\bn$ 
belonging to sublattice $B$ must have the same value, denoted as $\la_B$,
where $\la_A \la_B = 1$. Eq. (\ref{ecl}) then implies that $E_{cl} = 
(JSN/2) \la_A = (JSN/2) \la_B$. The condition $\la_A \la_B = 1$ then
implies that the minimum energy will be attained if $\la_A = \la_B = -1$.
Thus the classical ground state is equal to $-JSN/2$ corresponding
to $\la_{\bn} = -1$ at all sites.

We will now explicitly find a large set of classical ground states. To this 
end, we observe the following interesting {\it one to many} correspondence 
of dimer coverings on a honeycomb lattice with a set of classical ground states
of the Kitaev model that {\it have identical energy}. Consider a covering 
of the honeycomb lattice with dimers, such that every site lies on a dimer. 
Associate a classical spin configuration to each dimer (bond) 
as follows. Depending on whether it is an $x$, $y$ or $z$ bond, we put the 
two spins at the ends of the dimer as antiparallel and along the $\hat \bx$, 
$\hat \by$ or $\hat \bz$ direction respectively in spin space. This is the 
reason we call them Cartesian or CN-ground states. All these classical states 
have an identical energy $-JSN/2$. This follows from the fact that the 
interaction energy of the two spins of any dimer is $-JS$. Two neighboring
spins not belonging to a dimer have zero interaction energy, either because 
they are orthogonal or the corresponding spin components do not appear in 
the bond interaction term.

The number of dimer coverings on a honeycomb lattice has an asymptotic form 
$(1.381)^{N/2}$ \cite{baxter}. Further, the spins of each dimer can be in 
two possible antiparallel states; hence we have $2^{N/2}$ classical spin 
configurations for each dimer covering. This makes the total degeneracy of 
the CN-ground states to be $(1.662)^N$. 

An important question is whether these classical states remain stable
under quantum fluctuations. We answer this question in
two steps. First we show that the discrete and exponentially large set of
degenerate states found above are further connected by flat valleys in
the $N$-spin space. Then we perform a spin wave analysis and show that to
the leading order in $1/S$ there are no negative energy spin wave
excitations. This ensures local stability of our quantum ground states. 

The discrete set of degenerate states obtained from the dimer coverings forms 
a set of isolated points in the $N$-spin space, i.e., $\{S^2\}^N$, where $S^2$
denotes the surface of a sphere in three dimensions.
We will now show that there are flat valleys defined by a set of continuous
parameter which connect the discrete points. To see this consider a set of 
self avoiding walks (SAWs) that completely covers the lattice, such that each 
lattice site appears on one and only one SAW. In each SAW, let alternate bonds
form dimers. Each SAW must be either infinitely long or must be a closed loop 
consisting of an even number of bonds (this is because each bond on the 
honeycomb lattice goes from a site on sublattice $A$ to a site on sublattice 
$B$). 

Let us now consider the sites and bonds lying on one particular SAW. To be
specific, let us suppose that somewhere in the middle of the SAW, we
have some sites and bonds of the form $\cdots 1-x-2-z-3-y-4-x-5 \cdots$.
A discrete classical ground state is then given by one in which the
spins at the sites $1,2,3,4,5$ point along $\hat \bz, -\hat \bz, \hat
\bz, -\hat \bx, \hat \bx$; this has an energy of $-2JS$ for the four
bonds $xzyx$. Another discrete classical ground state is given by taking
the same five spins to point along $\hat \bx, -\hat \bx, \hat \by, -\hat
\by, \hat \by$. We now note that a continuous family of classical ground
states which interpolates between the above two discrete states is given
by a configuration in which the five spins point along $\sin \theta \hat
\bx + \cos \theta \hat \bz, - \sin \theta \hat \bx - \cos \theta \hat
\bz, \sin \theta \hat \by + \cos \theta \hat \bz, -\cos \theta \hat \bx
- \sin \theta \hat \by, \cos \theta \hat \bx + \sin \theta \hat \by$,
where $\theta$ goes from 0 to $\pi /2$; the energy of the four bonds for 
this configuration is $-2JS$ for all values of $\theta$. This transformation 
can be extended to all the sites of the SAW. 

Thus we have a continuous transformation taking us from one discrete 
classical ground state of a SAW to another; we will call such a transformation
a slide. A slide is parametrized by an angle $\theta$, and it takes us from 
a discrete classical ground state in which the even numbered bond energies 
are minimized to one in which the odd numbered bond energies are minimized. 
We now observe that if a slide rotates the spin on a particular
site in the $\hat \bx$ -
$\hat \by$ plane (such as site 4 in the previous paragraph), then that site 
must be coupled to a site in the neighboring SAW by a $z$ bond; hence the 
classical interaction energy of that site to its neighboring site on the 
other SAW remains zero throughout the slide. This is true for any two 
neighboring sites belonging to different SAWs, no matter in which plane 
each of them is rotated during the slides of the two SAWs. Thus,
a slide can be carried out on each SAW separately without changing
the classical interaction energy between the two SAWs; hence there is a 
continuous family of classical ground states on each SAW. However, based on 
the results in Sec. II, we expect that the zero point energy of the spin waves
about such a continuous family of ground states in a SAW will be minimized 
for a discrete set of values of $\theta$ which corresponds to the spin 
at each site pointing along one of the six directions $\pm \hat \bx$, $\pm 
\hat \by$ or $\pm \hat \bz$. We will therefore consider below only the 
discrete set of classical ground states described in the previous sentence.

We have seen that the discrete classical ground states correspond to dimer 
coverings of the honeycomb lattice. An interesting question to ask is whether 
all dimer coverings can be continuously connected to each other through
the continuous families of classical ground states. We will prove that the
answer is yes, by showing that any discrete classical ground state can be 
transformed by a succession of slides to a classical ground state in which 
all the dimers lie on the $z$ bonds (we will call these vertical dimers). For 
any dimer covering, the lattice can be covered by SAWs. We now choose these 
SAWs as follows. If all the dimers are vertical, there is nothing more to be
done. If at least one dimer is non-vertical, we consider that dimer;
each end of it also belongs to a vertical bond which is not a dimer
(since no point can belong to two dimers). We go to the other end
of that vertical bond; that end must belong to a non-vertical dimer. 
Continuing in this way, we get a SAW that consists of alternating 
non-vertical dimers and vertical bonds which are not dimers. We now
apply a slide to this SAW; we then get a SAW all of whose dimers are 
vertical. (Figure 2 shows an example of two SAWs which are connected 
by a slide). We then repeat the process of taking another non-vertical
dimer (which has no points in common with the previous SAW), constructing 
a new SAW from it using the above procedure, and finally performing a slide 
which converts all the dimers to vertical ones. By repeating this until
we have SAWs covering all the sites, we reach the state in which all
the dimers are vertical. Thus all dimer coverings of the honeycomb lattice
are connected through continuous families of classical ground states
to the purely vertical dimer covering. Therefore all dimer coverings are 
also connected to each other through continuous families. 

\begin{figure}[htb] \rotatebox{0}{\includegraphics*[width=\linewidth]{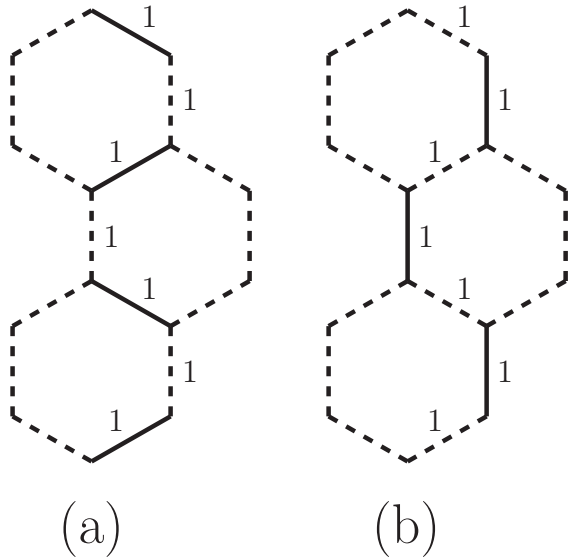}}
\caption{Pictures of two self avoiding walks, marked as 1, which are related 
to each other by a slide. The walk in (a) has only non-vertical dimers, while
the walk in (b) has only vertical dimers. The dimers are shown by solid 
lines.} \label{fig2} \end{figure}

We will now use the HP transformation to compute the spin wave spectrum about 
any one of the discrete classical ground states given by a dimer covering.
Consider a vertical dimer on which the two spins point along the $\pm \hat 
\bz$ direction. Let us consider these two spins separately.

\noi (i) For the spin pointing along the $\hat \bz$ direction, we have
$S^z = S - (p^2 + q^2)/2$, but $(S^x,S^y)$ can be chosen in four different
ways, namely, $\sqrt{S} (q,p), \sqrt{S} (-q,-p), \sqrt{S} (p,-q)$, and
$\sqrt{S} (-p,q)$, up to the lowest order in the HP transformation.

\noi (ii) For the spin pointing along the $-\hat \bz$ direction, we have $S^z
= - S + (p^2 + q^2)/2$, but $(S^x,S^y)$ can be chosen in four ways, namely, 
$\sqrt{S} (p,q), \sqrt{S} (-p,-q), \sqrt{S} (q,-p)$ and $\sqrt{S} (-q,p)$.

The term coupling the two spins at the opposite ends of the vertical
dimer is given by $S_i^z S_j^z = [S - (p_i^2 + q_i^2)/2] ~[-S + (p_j^2 + 
q_j^2)/2] \simeq - S^2 + (S/2) (p_i^2 + q_i^2 + p_j^2 + q_j^2)$ up to order 
$S$. Thus there is no coupling between the two spins to this order in $S$.

\begin{figure}[htb] \rotatebox{0}{\includegraphics*[width=\linewidth]{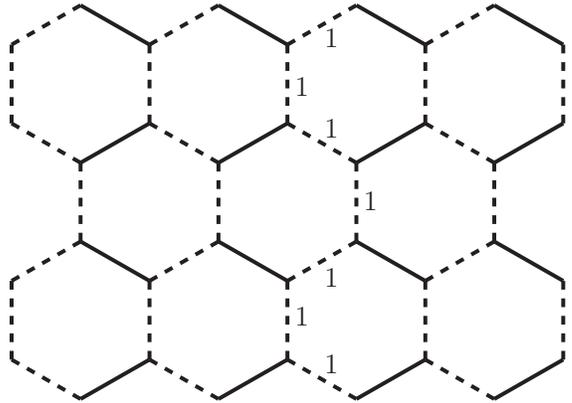}}
\caption{Picture of a set of self avoiding walks (dotted lines); one of the
walks is marked as 1. The dimers are shown by solid lines.} \label{fig3} 
\end{figure}

For any dimer covering of the honeycomb lattice, there is a set of self 
avoiding walks (SAW) covering the lattice such that none of the bonds 
appearing on a SAW is a dimer. Hence the two spins appearing at the two ends 
of any dimer belong to different SAWs, as shown in Fig. 3. (Notice that these 
SAWs are different from the ones we discussed earlier while finding our flat 
valley through the slide operation; in those SAWs, alternate bonds were 
dimers). As we saw above, sites 
belonging to different SAWs are decoupled from each other up to order $S$. 
We can therefore carry out a spin wave analysis for each SAW separately.
Now consider a SAW which forms a closed loop with $n$ sites, where we saw 
earlier that $n$ must be an even integer. (The minimum value of $n$ is 6 
corresponding to a hexagon). Let $n=2m$. As we go around the loop, we choose 
the spin variables along the loops to be $q$ and $p$ alternately, so that the 
couplings between nearest neighbors involve either $q_m,q_n$ or $p_m,p_n$, 
but not $q_m,p_n$. Because of the two cases (i) and (ii) discussed
above, the loop may have either periodic boundary condition (PBC) or 
antiperiodic boundary condition (ABC). Ignoring a constant, 
we find the spin wave Hamiltonian for the SAW to be
\bea H_{sw} &=& \frac{J}{2} ~\sum_{i=1}^n ~(p_i^2 ~+~ q_i^2)~ \non \\
& & +~ J~ \sum_{i=1}^{m-1} ~(p_{2i-1} p_{2i} ~+~ q_{2i} q_{2i+1}) \non \\
& & +~ J~ (p_{n-1} p_n ~\pm~ q_n q_1), \label{ham3} \eea
with either PBC or ABC for the last bond connecting sites $n$ and $1$; the 
sign of the $q_n q_1$ term is $+$ and $-$ in the two cases respectively. The 
SAW has $m$ unit cells, each consisting of 2 sites. The normal modes can be 
characterized by a momentum $k$, where $k = 0, ~2\pi /m, ~\cdots , ~(2\pi m - 
2\pi)/m$ in the case of PBC, and $k = \pi /m, ~3\pi /m, ~\cdots , ~(2\pi m - 
\pi)/m$ in the case of ABC. We now find the normal mode frequencies by solving
the classical Hamiltonian equations of motion. For each momentum $k$, we find 
that there are two frequencies given by 0 and $\omega_k = 2 J |\cos (k/2)|$.
The existence of a zero energy mode for each $k$ is a signature of the enormous
ground state degeneracy at the classical level; the zero mode also implies 
that the spin wave correction to the expectation value of the spin at each 
site diverges.

The spin wave normal modes are interesting. Even though they are formally 
characterized by a wave vector, they are defined on self avoiding strings of
varying shapes and sizes on the lattice. Secondly, by construction, the 
classical interaction energy of the spins on a SAW is identically zero, bond 
by bond; it is neither a minimum nor a maximum. We have a collection of 
{\it one-dimensional} spin waves living on SAWs. Further these spin waves
have a linear dispersion, $\omega_{\pi + q} \approx 2 J |q| $ at low 
frequencies, around $k = \pi$. 

A linear dispersion is known in the spin-1/2 Heisenberg antiferromagnetic 
system in one dimension. However, the spin waves there are spin-1 excitations.
In the Kitaev model under consideration, this linear dispersion occurs for both
antiferromagnetic and ferromagnetic couplings and is non-degenerate, indicating
some kind of {\it spin-zero} or {\it real scalar} character
of the spin wave quanta. The frustration in the Kitaev model seems to induce 
an effective antiferromagnetic behavior along the SAW lines whatever be the 
sign of $J$. This linear spin wave spectrum should be considered as a 
precursor to the linear Majorana spectrum that one gets for the spin-1/2 
Kitaev model. It is likely that these scalar spin wave quanta undergo quantum 
number fractionization leading to Majorana fermions.

% We call them {\it stringons}. 

We now calculate the zero point energy per site $e_0 =\sum_k \omega_k /(2n)$
as a function of $n \ge 6$. We find that

\noi (i) for $m$ odd, $e_0 /J = {\rm cosec} (\pi /n)/n$ for PBC and $\cot 
(\pi /n)/n$ for ABC.

\noi (ii) for $m$ even, $e_0 /J = \cot (\pi /n)/n$ for PBC and ${\rm cosec} 
(\pi /n)/n$ for ABC.

\noi In all cases, $e_0 /J \to 1/\pi \simeq 0.318$ as $n \to \infty$.
We find that the minimum value of $e_0$ occurs if $n=6$ and we have ABC. In 
that case, $e_0 /J = \sqrt{3}/6 \simeq 0.289$.

Now we show that the above minimum for the zero point energy per site can 
actually be achieved for the entire honeycomb lattice. Consider a dimer
covering such that the corresponding SAWs cover the lattice with hexagons,
as shown in Fig. 4.
(There are three such dimer coverings; this will contribute a factor of 3
when we compute the ground state degeneracy below). For each such dimer
covering, we take the two spins on all dimers on bonds of type $n$ to 
point in the same way, say, the top spin pointing along $\hat \bn$ and the 
bottom spin pointing along $- \hat \bn$ (here $n = x, y$ or $z$). Then 
we find that each hexagon has ABC and therefore $e_0 = 0.289J$.

\begin{figure}[htb] \rotatebox{0}{\includegraphics*[width=\linewidth]{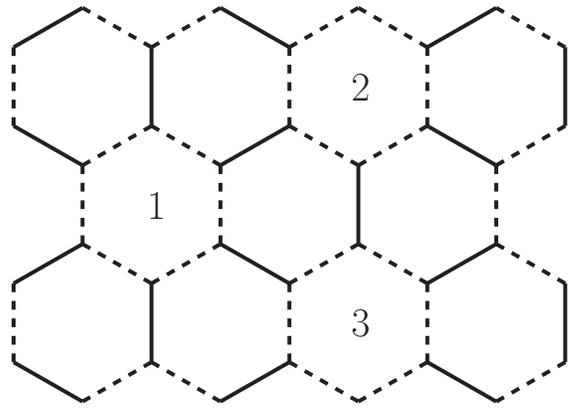}}
\caption{Picture of a ground state in which the self avoiding walks (dotted 
lines) form hexagons; three of these are marked as 1-3. The dimers are shown 
by solid lines.} \label{fig4} \end{figure}

Next, we note that for each of the three dimer coverings, the SAWs cover the 
lattice with $N/6$ hexagons. These hexagons form a triangular lattice 
containing $N/3$ triangles. We then see that the ABC of the Hamiltonian
in Eq. (\ref{ham3}) continues to hold if we flip both the spins on all 
the three dimers which pairwise join three hexagons which form a triangle.
So even after the dimer covering is fixed, we still have an exponentially 
large number of ground states given by $2^{N/3}$ corresponding to 
the different ways in which pairs of spins on different dimers can point.
Since there are three possible dimer coverings, the total degeneracy of
the quantum ground states is $3 \times 2^{N/3}$ which goes as $(1.260)^N$
for large $N$. This is a smaller exponential than the number of discrete
classical ground states which goes as $(1.662)^N$ as stated earlier.

To conclude, we have shown that the dimer coverings which have the minimum 
zero point energy per site given by $0.289J$ are the ones which correspond
to SAWs covering the lattice with hexagons. The number of quantum ground
states picked out by the zero point energy of the spin waves still grows
exponentially with the number of sites, but it grows more slowly than the
number of classical ground states.

Finally, we can again consider what happens if the couplings on the
three kinds of bonds are different, say, $J_x$, $J_y$ and $J_z$. We again
find that if one of these, say $J_z$, is larger than the other two, then
the classical ground states are given by the state with purely vertical 
dimers; in each such dimer, the two spins can point along the $+\hat \bz, - 
\hat \bz$ or $-\hat \bz, + \hat \bz$ directions. Hence the number of classical
ground states is $2^{N/2}$. We find that this degeneracy is not broken by the 
zero point energy of the spin waves.

\section{Conserved $Z_2$ fluxes}

In this section we construct commuting operators for the spin-$S$ Kitaev model
and also generalize the Jordan-Wigner transformation for the spin-$S$ case. 
In the process we get exact results and some new insights.

For the spin-1/2 Kitaev model in two dimensions, it is known that there is a 
conserved quantum number associated with each hexagon. When the model is
rewritten as a $Z_2$ gauge theory, these conserved quantities correspond
to the $Z_2$ fluxes passing through the hexagon. Since the number of such 
quantum numbers (or hexagons) is $N/2$, the Hilbert space decomposes into 
$2^{N/2}$ independent sectors corresponding to each flux independently taking 
the values $\pm 1$. We will now show that all this continues to hold for 
arbitrary values of the spin $S$, integer or half-odd-integer.

We first note that the three spin operators $S^x, S^y, S^z$ satisfy the 
identities
\bea e^{i\pi S^a} ~S^b ~e^{-i\pi S^a} &=& S^b ~~{\rm if} ~~a ~=~ b, \non \\
&=& - ~S^b ~~{\rm if} ~~a ~\ne~ b, \label{id1} \eea
for $a,b=x,y,z$. Now consider a hexagon $h$ with sites labeled 
$1, ~\cdots, ~6$, with the Hamiltonian
\beq H_h ~=~ S_1^x S_2^x ~+~ S_2^y S_3^y ~+~ S_3^z S_4^z ~+~ S_4^x S_5^x ~+~ 
S_5^y S_6^y ~+~ S_6^z S_1^z. \eeq
If we define an operator
\beq W_p ~\equiv ~e^{i \pi (S_1^y ~+~ S_2^z ~+~ S_3^x ~+~ S_4^y ~+~ S_5^z~
+~ S_6^x)} \label{fluxop} \eeq
(see Fig. 1), then it follows from Eq. (\ref{id1}) that
\beq W_p H_h (W_p)^{-1} ~=~ H_h. \label{ocomm} \eeq
Thus $W_p$ commutes with the Hamiltonian $H_h$. It is easy to check that 
$W_p$ also commutes with the other terms of the full Hamiltonian coming
from other plaquettes. For $S=1/2$, we observe that
\beq W_p ~=~ -~ \si_1^y ~\si_2^z ~\si_3^x ~\si_4^y ~ \si_5^z ~\si_6^x , 
\label{fluxop12} \eeq
where $\si^a$ denote the Pauli matrices.

Since $e^{\pm i 2\pi S^a}=(-1)^{2S}$, we have $e^{i\pi S^a}=(-1)^{2S}
e^{-i\pi S^a}$. It then follows from Eq. (\ref{id1}) that
\beq e^{i\pi S^a} e^{i\pi S^b} ~=~ (-1)^{2S} e^{i\pi S^b}e^{i\pi S^a} ~~
{\rm if} ~~a ~\ne~ b. \label{id2} \eeq
It then follows that $W_p$ will commute with $W_{p^\prime}$ since the 
two will share an even number of sites. It also follows that $W_p^2 = 1$. 
Thus the $W_p$'s are a set of mutually commuting conserved operators with 
eigenvalues equal to $\pm 1$. 

We note that for $S=1/2$, these operators are the same as the conserved
flux operators. We will henceforth refer to these operators as the flux
operators for all $S$. While we do not have a gauge theoretic formalism 
of the model for $S > 1/2$, we note that we can associate a conserved $Z_2$ 
quantum number with every closed loop on the lattice as follows. For every 
site on the closed loop, we define the normal direction as the direction 
associated with the bond that does not belong to the loop. If 
$(i_1,i_2,\dots,i_N)$ are the sites for that loop and $(a_1,a_2,\dots,a_N)$ 
are the corresponding normal directions, then the conserved quantity is
\beq W_L ~=~ \prod_{n=1}^N ~e^{i\pi S_{i_n}^{a_n}} ~. \label{loopcons} \eeq
It is interesting to note that these closed string operators are very similar 
to those defined by Den Nijs and Rommelse \cite{nijs} in the context of $S=1$
chains and generalized to arbitrary values of $S$ by Oshikawa \cite{oshikawa}.

\subsection{The ground state fluxes}

In the $S=1/2$ model, it has been proven that the values of the flux operators
are the same for all the elementary hexagonal plaquettes and are equal to $+1$.
Our semiclassical results indicate that this may be true for all values of $S$.

Consider the quantum state constructed by taking the direct product of
the spin coherent states corresponding to any of the classical ground state 
configurations defined by a dimer covering of the lattice. We will refer
to these states as the semiclassical ground states. Consider a 
dimer covering which defines a set of closed self avoiding loops 
such that the dimers form the normal directions to these loops. The 
semiclassical ground state corresponding to such a dimer covering is a 
simultaneous eigenstate of the flux operators corresponding to the 
closed loops defined by the dimer covering. The eigenvalues are given by
\beq \prod_{n=1}^N ~e^{i\pi S_{i_n}^{a_n}} ~\vert\psi\rangle ~=~
e^{i\pi S\sum_{n=1}^N~p_n} ~\vert\psi\rangle , \label{fluxev} \eeq
where $p_n=\pm 1$ depending on the polarization of the spin at the
$n^{\rm th}$ site. Since the honeycomb lattice is bipartite, all closed loops 
have an even number of sites. The eigenvalues are thus $e^{i 2\pi lS}$, where 
$l$ is an integer. If $S$ is an integer, then we see that the eigenvalues are 
always 1. If $S$ is a half-odd-integer, then it is $(-1)^l$. It can be checked
that the classical configurations which lead to ABC in the spin wave 
Hamiltonian correspond to even $l$.

Thus we see that the spin wave fluctuations pick out the states which are 
simultaneous eigenstates of the flux operators corresponding to one third of 
the elementary plaquettes. It further picks out the states which have 
eigenvalue $+1$ for these operators. We note that the semiclassical ground 
states can never be the simultaneous eigenstate of all the flux operators. 
The spin wave fluctuations seem to pick out the states which are the 
simultaneous eigenstates of the maximal number of elementary plaquette flux 
operators with eigenvalue $+1$. In the $S=1/2$ case, the exact ground state 
is a simultaneous eigenstate of all the flux operators with eigenvalue $+1$. 
Our results indicate that this may be the case for all $S$ also.

\subsection{The flux basis and spin-spin correlations}

One of the intriguing features of the Kitaev model is the peculiar form
of the spin-spin correlation functions \cite{baskaran1}. Only the nearest
neighbor correlations are non-zero. Further, only $S^x_iS^x_j$ on the 
$x$ bonds, $S^y_iS^y_j$ on the $y$ bonds and $S^z_iS^z_j$ on the $z$ bonds 
have non-zero values. These results are true for not only the ground
state but also for any eigenstate of the model. In fact, as we will show,
this form of the correlation function characterizes the set of simultaneous 
eigenstates of the flux operators.

Consider any simultaneous eigenstate of all the elementary flux
operators denoted by $\vert \{p_n\}\rangle$, where $p_n=\pm 1$
is the eigenvalue of the flux operator of the $n^{\rm th}$ plaquette.
The spin operator $S^a_i$ acting on this state produces another
simultaneous eigenstate of the flux operators, with the eigenvalues
of the two plaquettes which share the bond $(i,i+\hat a)$ flipped.
This follows from Eqs. (\ref{id2}) and (\ref{fluxop}). As two states 
with different sets of values of $p_n$ are orthogonal, the only non-zero 
spin-spin correlations are
\beq \langle S^x_iS^x_{i+\hat x}\rangle,~~ \langle S^y_iS^y_{i+\hat y}
\rangle,~~ {\rm and} ~~\langle S^z_iS^z_{i+\hat z} \rangle . \label{nzcorr} 
\eeq

Thus, as stated earlier, the flux operators define a basis of a peculiar 
kind of spin liquid. This basis is formally constructed in the fermionic
formalism of the spin-1/2 model. We will now show that it is possible to
have an analogous Jordan-Wigner construction to construct this basis in
the spin-$S$ model also. However, unlike the spin-1/2 case our
construction will not exhaust all the states in the Hilbert space and
thus does not lead to an exact solution of the model.

Consider a Hamilton path running through the lattice. For simplicity, we 
consider an infinite lattice and take the path to go through the $x$ and $y$ 
bonds only. The construction can be generalized to any SAW \cite{baskaran2}.
Define the ``disorder operators" on the $n^{\rm th}$ site to be
\beq \mu_{i_n} ~=~ \prod_{m<n} ~e^{i\pi (S^z_{i_m}+S)}~, \label{disopdef} \eeq
where the $\mu_{i_n}$'s commute with each other and $\mu_{i_n}^2=1$.
At each site $i_n$, the Hamilton path is used to classify the three
bonds as {\em incoming, outgoing} and {\em normal}. The normal bond for
our path is the $z$ bond. We denote the spin operator in the incoming
and outgoing bond directions by $S^{t_1}_{i_n}$ and $S^{t_2}_{i_n}$
respectively. We now define two operators at each site,
\bea \xi_{i_n}&\equiv& e^{i\pi (S_{i_n}^{t_1}+S)}\mu_{i_n}, \non \\
\chi_{i_n}&\equiv& e^{i\pi (S_{i_n}^{t_2}+S)}\mu_{i_n}. \label{xidef} \eea
It then follows that
\bea \xi_{i_n}\xi_{i_m}-(-1)^{2S}\xi_{i_m}\xi_{i_n} &=& \delta_{nm}, \non \\
\chi_{i_n}\chi_{i_m}-(-1)^{2S}\chi_{i_m}\chi_{i_n} &=& \delta_{nm}, \non \\
\xi_{i_n}\chi_{i_m}-(-1)^{2S}\chi_{i_m}\xi_{i_n} &=& 0. \eea
Thus the $\xi$ and $\chi$ operators are Majorana fermions for 
half-odd-integer spins and hard-core bosons for integer spins.

We now consider the commutators of the $\xi_{i_n}$ and $\chi_{i_n}$ 
with the Hamiltonian. Consider the terms $\xi_{i_n}H\xi_{i_n}$ and
$\chi_{i_n}H\chi_{i_n}$. All the terms in the Hamiltonian involving spins at 
sites $i_m$ for $m<n$ are left invariant. This is because either both or none
of the spins have their signs flipped. So we have
\bea \xi_{i_n}H\xi_{i_n}&=&H-2J_{t_2} S_{i_n}^{t_2}S_{i_{n-1}}^{t_2}
-2J_{a_n} S_{i_n}^{a_n}S_{i_n+\hat a_n}^{a_n}, \non \\
\chi_{i_n}H\chi_{i_n}&=& H-2J_{a_n}S_{i_n}^{a_n}S_{i_n+\hat a_n}^{a_n}. \eea
It then follows that the operators defined on the normal bonds,
\beq u_{ij}=e^{i\pi S}\chi_i\chi_j, \eeq
form a set of mutually commuting operators which also commute with the 
Hamiltonian. The flux operator on any elementary plaquette is equal to
the product of the $u_{ij}$ operators on the two $z$ bonds of the
hexagon just as in the $S=1/2$ case.

Thus, just as in the $S=1/2$ case, the flux basis is easy to construct in 
terms of the $\chi_i$ operators. However, writing it in a simple form in 
terms of the original spins remains a challenge.

\section{An Exactly solvable Higher spin Model with Free Majorana Fermions}

For the case of half-odd-integer spin $S$, we find that it is not possible 
to write the spin-$S$ Kitaev model in Eq. (\ref{ham2}) in terms of local 
interactions between the Majorana fermions introduced above. The difficulty 
arises from not being able to invert Eq. (\ref{xidef}) to obtain the spin 
operators. However, for any finite value of half-odd-integer spin, we can 
define a modified Kitaev Hamiltonian, for which we can give the exact spectrum
and degeneracies. Our modified Kitaev Hamiltonian has the form
\bea H &=& \frac{J}{S} ~\sum_{j+l={\rm even}} ~(\tau_{j,l}^x \tau_{j+1,l}^x~
+~ \tau_{j-1,l}^y \tau_{j,l}^y \non \\
& & ~~~~~~~~~~~~~~~~+~ \tau_{j,l}^z \tau_{j,l+1}^z), \label{ham4} \eea
where $\tau^a \equiv e^{i\pi S^{a}}$ is the $\pi$-rotation operator 
introduced in Eq. (\ref{id1}).
% , which must reduce to a polynomial in $S^{a}$ of 
% degree $2S + 1$ because of the Cayley-Hamilton theorem.
Now, the operators $\tau^a$ only connect states which have the same magnitude 
of $S^z$, in the $S^z$ basis; this is because $e^{i \pi S^x}$ and $e^{i \pi 
S^y}$ acting on $| S_z = m \rangle$ give $| S_z = -m \rangle$. Hence the 
$\tau^a$'s reduce to $2 \times 2$ blocks in the basis of eigenstates of $S^z$;
the number of such blocks is equal to $S + 1/2$ corresponding to $m = 1/2, ~
3/2, ~\cdots, ~S$.
For example, in the case of $S = 3/2$, we find that in the basis of the two 
states with $S^z = \pm 3/2$, $\tau^x = -i \sigma^x$, $\tau^y = i \sigma^y$,
and $\tau^z = -i \sigma^z$, while in the basis of the two states with $S^z = 
\pm 1/2$, $\tau^x = - i \sigma^x$, $\tau^y = -i \sigma^y$, and $\tau^z = i 
\sigma^z$.

In view of this, the $(2S+1)^N$ dimensional Hilbert space for $N$ spins 
decomposes into $(S + 1/2)^N$ copies of $2^N$ dimensional Hilbert spaces. 
Inside each copy, the Hamiltonian in Eq. (\ref{ham4}) behaves exactly like 
the Kitaev Hamiltonian in Eq. (\ref{ham2}), leading to the identical spectrum 
and physical properties. The degeneracy of $(S + 1/2)^N$ of the Hamiltonian 
(which is not a gauge degeneracy) is an unusual decomposition of the 
eigenstates and may have some special use in quantum computation.

In the case of integer spins, Eq. (\ref{id2}) shows that the $\tau^a$ 
operators commute with each other for all the values of $a=x,y,z$. Hence, they
can be diagonalized simultaneously; each of the diagonal entries is equal to 
$\pm 1$. The integer spin version of Eq. (\ref{ham4}) therefore reduces to a 
kind of Ising spin model which is classical rather than quantum mechanical.

\section{Discussion}

In this paper we have presented a large spin analysis of the Kitaev
model, whose spin-1/2 end is exactly solvable. We find a classical
ground state structure which has a non-trivial geometry in the $N$-spin
space. There are discrete sets of points that are connected by flat
valleys. Our spin wave analysis gives either zero or positive energy
excitations, indicating local stability of the degenerate set of vacuua.
Further, depending on the vacuum chosen (which depends on the dimer covering 
pattern), the spin waves are localized on the SAW curves. It would be 
interesting to see how and when quantum number fractionization occurs and 
Majorana fermions emerge, when one goes beyond harmonic spin wave theory.

A class of excitations that we have not studied in this paper is the one
involving dimer coverings containing one or more defects. By definition, a 
defect site does not lie on a dimer. We can consider ground states about 
these defective dimer coverings; these will define ground states containing 
a topological spin defect. As the energy of one topological defect is of 
the order of $JS$, there is a finite gap to these excitations. It would be 
interesting to study whether such a defect might be related to a Majorana 
fermion excitation.

\section*{Acknowledgments}

D.S. thanks DST of India for financial support under Project No.
SR/S2/CMP-27/2006.

\end{document}